\documentclass[aps,floatfix,showpacs,superscriptaddress,showkeys]{revtex4}
\usepackage{amssymb,amsmath,amsfonts,latexsym,graphicx,epsfig,here}
\newcommand{\bea}{\begin{eqnarray}}
\newcommand{\ena}{\end{eqnarray}}
\newcommand{\be}{\begin{equation}}
\newcommand{\en}{\end{equation}}
\newcommand{\nn}{\nonumber\\}

\newcommand{\la}{\langle}
\newcommand{\ra}{\rangle}

\begin{document}

\hfill MITP/17-051 (Mainz)

\title{Decay chain information on the newly discovered double charm baryon 
state $\Xi_{cc}^{++}$} 

\author{Thomas Gutsche}
\affiliation{Institut f\"ur Theoretische Physik, Universit\"at T\"ubingen,
Kepler Center for Astro and Particle Physics, 
Auf der Morgenstelle 14, D-72076 T\"ubingen, Germany}
\author{Mikhail~A.~Ivanov}
\affiliation{Bogoliubov Laboratory of Theoretical Physics,
Joint Institute for Nuclear Research, 141980 Dubna, Russia}
\author{J\"urgen~G.~K\"orner}
\affiliation{PRISMA Cluster of Excellence, Institut f\"{u}r Physik,
Johannes Gutenberg-Universit\"{a}t, D-55099 Mainz, Germany}
\author{Valery E. Lyubovitskij}
\affiliation{Institut f\"ur Theoretische Physik, Universit\"at T\"ubingen,
Kepler Center for Astro and Particle Physics,
Auf der Morgenstelle 14, D-72076 T\"ubingen, Germany}
\affiliation{Departamento de F\'\i sica y Centro Cient\'\i fico
Tecnol\'ogico de Valpara\'\i so-CCTVal, Universidad T\'ecnica
Federico Santa Mar\'\i a, Casilla 110-V, Valpara\'\i so, Chile}
\affiliation{Department of Physics, Tomsk State University,
634050 Tomsk, Russia} 
\affiliation{Laboratory of Particle Physics,
Tomsk Polytechnic University, 634050 Tomsk, Russia}

\begin{abstract}

  We interprete the new double charm baryon state found by the LHCb
  Collaboration in the invariant
  mass distribution of the set of final state particles
  $(\Lambda_c^+\,K^-\,\pi^+\,\pi^+)$ as being at the origin of the
  decay chain
  $\Xi_{cc}^{++} \to \Sigma_c^{++} (\to \Lambda_c^+ \pi^+)
  +\bar K^{*0} (\to K^-  \pi^+)$. The nonleptonic decay
  $\Xi_{cc}^{++} \to \Sigma_c^{++} + \bar K^{*0}$ belongs 
  to a class of decays
  where the quark flavor composition is such that the decay proceeds solely
  via the factorizing contribution precluding a contamination from internal
  $W$-exchange. We use the covariant confined quark model previously
  developed by us to calculate the four helicity amplitudes that
  describe the dynamics of the transition $\Xi_{cc}^{++} \to \Sigma_c^{++}$ 
  induced by the effective $(c \to u)$ current. We then proceed to  
   calculate the rate of the decay as well as the polarization of the
   $\Sigma_c^{++}$ and $\Lambda_c^+$ baryons and the
   longitudinal/transverse composition of the $\bar K^{*0}$. 
   We estimate the decay $\Xi_{cc}^{++} \to \Sigma_c^{++} \bar K^{*0}$ 
   to have a branching rate of 
   $B(\Xi_{cc}^{++} \to \Sigma_c^{++} \bar K^{*0}) \sim 10.5 \%$. 
   As a byproduct of our investigation we have also analyzed the decay 
   $\Xi_{cc}^{++} \to \Sigma_c^{++} \bar K^{0}$ for which we find
   a branching ratio of 
   $B(\Xi_{cc}^{++} \to \Sigma_c^{++} \bar K^0) \sim 2.5 \%$. 

\end{abstract}

\today


\maketitle

\section{Introduction}

Very recently the LHCb Collaboration has reported on the discovery of the
double charm state $\Xi_{cc}^{++}$~\cite{Aaij:2017ueg} found in the invariant 
mass spectrum of the final state particles $(\Lambda_c^+\,K^-\,\pi^+\,\pi^+)$
where the $\Lambda_c^+$ baryon was reconstructed in the decay mode
$pK^-\pi^+$. The mass of the new state was given as 
$3621.40 \pm 0.72 \pm 0.27 \pm 0.14$ MeV. 
The central value of the extracted mass 
is very close to the 3610 MeV value predicted 
in Ref.~\cite{Korner:1994nh} in the framework of the one gluon exchange model
of de Rujula, Georgi and Glashow~\cite{DeRujula:1975smg} which features 
a Breit-Fermi spin-spin interaction term. It is noteworthy
that Ebert et al. predicted a mass of 3620 MeV for the $\Xi_{cc}^{++}$ using 
a relativistic quark-diquark potential model~\cite{Ebert:2002ig}.
We interprete the new double charm baryon state found in the
$(\Lambda_c^+\,K^-\,\pi^+\,\pi^+)$ mass distribution
as being at the origin of the decay chain 
  $\Xi_{cc}^{++} \to \Sigma_c^{++} (\to \Lambda_c^+ \pi^+)
+ \bar K^{*0} (\to K^-  \pi^+)$. This decay chain is favored from
an experimental point of view since the branching ratios of the daughter
particle decays $\Sigma_c^{++} \to \Lambda_c^+ \pi^+$ and
$ \bar K^{*0} \to K^-  \pi^+$ are large ($\sim 100\,\%$ and, from isospin
invariance, $\sim 66\,\%$, respectively).
       
The nonleptonic decay
  $\Xi_{cc}^{++} \to \Sigma_c^{++} +\bar K^{*0}$ belongs to a class of decays
  where the quark flavor composition is such that the decay proceeds solely
  via the factorizing contribution precluding a contamination from internal
  $W$-exchange. We use the covariant confined quark model previously proposed 
  and developed by us to calculate the four helicity amplitudes that
  describe the dynamics of the transition
  $\Xi_{cc}^{++} \to \Sigma_c^{++}$ induced by the effective $(c \to u)$ 
  current.
  We then proceed to calculate the rate of the decay as well as the 
  polarization of the $\Sigma_c^{++}$ and $\Lambda_c^+$ baryons and the
   longitudinal/transverse
   composition of the $\bar K^{*0}$. The nontrivial helicity composition of
   the $\bar K^{*0}$ leads to a nontrivial angular decay distribution in
   terms of
   the polar angle $\theta_V$ formed by the direction of the $K^-$ in the
   $\bar K^{*0}$ rest system and the original
   flight direction of the $\bar K^{*0}$.

   Double heavy baryon decays and their magnetic moments were
   treated by us before in Refs.~\cite{Faessler:2001mr} where we 
   performed a comprehensive study of the semileptonic and radiative decays 
   of double heavy baryons using a covariant quark model. 
   The version of the covariant quark
   model used in Ref.~\cite{Faessler:2001mr} has recently been improved by
   incorporating quark confinement in an effective way~\cite{Branz:2009cd}.
   For the calculation of the relevant transition
   $\Xi_{cc}^{++} \to \Sigma_c^{++}$ in this paper we use the improved model
   which we dub the covariant confined quark model (CCQM). 
   The physics of double heavy charm and bottom baryons (mass spectrum and
   decay properties) has been studied before in a number of 
   papers~\cite{Fleck:1989mb}-\cite{Chen:2017sbg}.

\section{Matrix elements, helicity amplitudes and rate expressions}

The matrix element of the exclusive decay 
$B_1(p_1,\lambda_1)\to B_2(p_2,\lambda_2)\,+\,V(q,\lambda_V)$ 
 is defined by
\be
M(B_1\to B_2 + V) =  
\frac{G_F}{\sqrt{2}} \, V_{ij} \, V^\ast_{kl} \, C_{\rm eff} \, 
f_V \, M_V \, \la B_2 | \bar q_2 O_\mu q_1 | B_1 \ra \, 
\epsilon^{\dagger\,\mu}(\lambda_V) \,.
\label{eq:matr_LbLV}
\en
In the present case $B_1=\Xi_{cc}^{++}$, $B_2=\Sigma_c^{++}$, 
$V=\bar K^{*\,0}$, $V_{ij} = V_{ud} = 0.97425$, 
$V^\ast_{kl} = V^\ast_{cs} = 0.974642$. 
For the effective current strength we use the large $N_c$ limit of 
the relevant effective current combination 
$C_{\rm eff} = C_2 + 1/N_c \,\cdot\, C_1$ to write
$C_{\rm eff} = - 0.565$~\cite{Buchalla:1995vs}. 
The large $N_c$ limit has also been used 
to successfully describe the nonleptonic decays  
$\Lambda_b^0 \to \Lambda + J/\psi$ and 
$\Lambda_c^+ \to p \phi$ which belong to same class of neutral vector meson
decays as $\Xi_{cc}^{++} \to \Sigma_c^{++} +\bar K^{*0}$ which
proceed solely via the factorizing contribution (also called internal
$W$-emission) in Refs.~\cite{Ivanov:1996fj,Gutsche:2013oea}. 
The leptonic decay constant is denoted by $f_V$. 
The Dirac string $O^\mu$ reads $O^\mu = \gamma^\mu (1 - \gamma^5)$.

The hadronic matrix element $\la B_2 | \bar q_2 O_\mu q_1 | B_1 \ra$
is expressed in terms of four dimensionless invariant form factors
$F^{V/A}_{1,2}(q^2)$, viz.
\bea 
\la B_2 | \bar q_2 \gamma_\mu q_1 | B_1 \ra &=& 
\bar u(p_2,s_2) 
\Big[ \gamma_\mu F_1^V(q^2) 
    - i \sigma_{\mu\nu} \frac{q_\nu}{M_1} F_2^V(q^2)       
\Big] 
u(p_1,s_1)\,, \nonumber\\[.3cm]
\la B_2 | \bar q_2 \gamma_\mu \gamma_5 q_1 | B_1 \ra &=& 
\bar u(p_2,s_2) 
\Big[ \gamma_\mu F_1^A(q^2) 
    - i \sigma_{\mu\nu} \frac{q_\nu}{M_1} F_2^A(q^2)       
\Big] 
\gamma_5 u(p_1,s_1)\,,
\ena 
where 
$\sigma_{\mu\nu} = (i/2) (\gamma_\mu \gamma_\nu - \gamma_\nu \gamma_\mu)$ 
and all $\gamma$-matrices are defined as in Bjorken-Drell. Here we 
drop $F^{V/A}_{3}(q^2)$ form factors, which do not contribute 
to the decay $B_1 \to B_2 + V$ due to the vector current conservation.

Next we express the vector and axial vector helicity amplitudes 
$H^{V/A}_{\lambda_2\lambda_V}$ contributing to the decay 
$\Xi_{cc}^{++} \to \Sigma_c^{++} + \bar K^{\star 0}$ 
in terms of the invariant form factors 
$F_{1,2}^{V/A}$,  where $\lambda_V = \pm 1, 0$ and 
$\lambda_2 = \pm 1/2$ are  the helicity components of the vector meson
and the baryon $B_2$,  respectively. 

We need to calculate the expression 
\be
H_{\lambda_2\lambda_V} = 
\la B_2(p_2,\lambda_2) |\bar q_2 O_\mu q_1 | B_1(p_1,\lambda_1) \ra
\, \epsilon^{\dagger\,\mu}(\lambda_V) 
=  H_{\lambda_2\lambda_V}^V - H_{\lambda_2\lambda_V}^A\,, 
\en 
where we split the  helicity amplitudes into their vector and axial parts. 
We shall work in the rest frame of the baryon $B_1$ with the baryon $B_2$ 
moving in the positive $z$-direction:
$p_1 = (M_1, \vec{\bf 0})$, $p_2 = (E_2, 0, 0, |{\bf p}_2|)$ and 
$q = (q_0, 0, 0, - |{\bf p}_2|)$. The helicities of the $B_1, B_2$, and 
$V$ are related by $\lambda_1 = \lambda_2 - \lambda_V$ leading to the angular
momentum restriction
$|\lambda_2 -\lambda_V|\le 1/2$.
One has
\bea
\begin{array}{lcrlcl}
H_{\frac12 0}^V &=& \sqrt{Q_-/M_V^2} \,
\Big( F_1^V M_+ + F_2^V \frac{M_V^2}{M_1} \Big)\,, \qquad 
& \qquad 
H_{\frac12 0}^A &=& \sqrt{Q_+/ M_V^2}\,   
\Big( F_1^A M_- - F_2^A \frac{M_V^2}{M_1} \Big)\,, 
\\[.3cm]
H_{\frac12 1}^V &=& \sqrt{2Q_-} 
\Big( - F_1^V - F_2^V \frac{M_+}{M_1} \Big)\,, \qquad 
& \qquad 
H_{\frac12 1}^A &=& \sqrt{2Q_+} \,  
\Big( - F_1^A + F_2^A \frac{M_-}{M_1} \Big)\,.  
\end{array}
\ena 
The remaining helicity amplitudes are given by the parity relations
$H^V_{-\lambda_2,-\lambda_V} = + H^V_{\lambda_2,\lambda_V}$, 
and $H^A_{-\lambda_2,-\lambda_V} = - H^A_{\lambda_2,\lambda_V}$.  
We use the abbreviations
$M_\pm = M_1 \pm M_2$, 
$Q_\pm = M_\pm^2 - M_V^2$,
${|\bf p_2|} = \lambda^{1/2}(M_1^2,M_2^2,M_{V}^2)/(2M_1)$. 

The decay width is given by 
\be
\Gamma(B_1 \to B_2\,+\,V) 
= \frac{G_F^2}{32 \pi} \, \frac{|{\bf p_2}|}{M_1^2} \, 
|V_{ij} V^\ast_{kl}|^2 \, C_{\rm eff}^2 \, f_V^2 \, M_V^2 \cdot 
\Big( {\cal H}_U + {\cal H}_L \Big) 
\label{eq:LbLV_width}\,, 
\en
where we introduce the following
combinations of helicity amplitudes
\bea
\qquad
\begin{array}{lr}
\hspace*{-.5cm}
\mbox{$ {\cal H}_U = |H_{\frac{1}{2}1}|^2 +  |H_{-\frac{1}{2}-1}|^2$} &
\hfill\mbox{ \rm transverse unpolarized}\,,
\\[.5cm]
\hspace*{-.5cm}
\mbox{$ {\cal H}_L = |H_{\frac{1}{2}0}|^2 +  |H_{-\frac{1}{2}0}|^2$} &
\hfill\mbox{ \rm longitudinal unpolarized}\,. 
\\
\end{array}
\ena 

   After having set up the spin-kinematical framework of the problem we now
   turn to the dynamics of the decay process  
   $\Xi_{cc}^{++} \to \Sigma_{c}^{++} \,\bar K^{*0}$ which necessarily 
   is model dependent. As remarked on before the sole contribution 
   to the nonleptonic decay 
   $\Xi_{cc}^{++} \to \Sigma_{c}^{++} \,\bar K^{*0}$   
   is the factorizing (or tree graph) contribution.  
   We use the CCQM to calculate the transition matrix 
   element $\Xi_{cc}^{++} \to \Sigma_{c}^{++}$.  

   An important ingredient of
   the calculation is the choice of the nonlocal interpolating current
   which we now specify together with the Lagrangian that describes the
   coupling of the constituent quarks with the double heavy baryon.
   One has~\cite{Faessler:2001mr,Ivanov:1996fj} 
\bea
\Xi_{cc}^{++}:\qquad &&{\cal L}^{\Xi_{cc}^{++}}_{\rm int}(x)
 = g_{\Xi_{cc}^{++}} \,\bar \Xi_{cc}^{++}(x)\cdot J_{\Xi_{cc}^{++}}(x) 
+ \mathrm{H.c.}\,,
\nonumber\\
&&J_{\Xi_{cc}^{++}}(x)
= \int\!\! dx_1 \!\! \int\!\! dx_2 \!\! \int\!\! dx_3 \,
F_{\Xi_{cc}^{++}}(x;x_1,x_2,x_3) \,
\epsilon^{a_1a_2a_3} \, \gamma^\mu \gamma^5 \, 
u^{a_1}(x_1)\,c^{a_2}(x_2) \,C \, \gamma_\mu \,
c^{a_3}(x_3)\,,
\label{eq:lag_Xicc}\\
\Sigma_{c}^{++}:\qquad &&{\cal L}^{\Sigma_c^{++}}_{\rm int}(x)
= g_{\Sigma_{c}^{++}} \,\bar \Sigma_{c}^{++}(x)\cdot J_{\Sigma_{c}^{++}}(x) 
+ \mathrm{H.c.}\,,
\nonumber\\
&&J_{\Sigma_{c}^{++}}(x)
= \int\!\! dx_1 \!\! \int\!\! dx_2 \!\! \int\!\! dx_3 \,
F_{\Sigma_{c}^{++}}(x;x_1,x_2,x_3) \,
 \epsilon^{a_1a_2a_3} \,\gamma^\mu \gamma^5 \,  
c^{a_1}(x_1)\,u^{a_2}(x_2) \,C \, \gamma_\mu \,
u^{a_3}(x_3)\,.
\label{eq:lag_Sigmac}
\ena
Differing from the calculations in~\cite{Ebert:2004ck,Wang:2017mqp}, 
which use a quark-diquark picture, we 
treat each of the three constituent quarks as separate dynamic entities. 
The propagators $S_q(k) = 1/(m_q - \not\! k)$ for up and charm quarks 
are taken in a form of free fermion propagators 
where $m_q = m_u, m_c$ are constituent quark masses fixed 
in previous analysis of a multitude of hadronic processes in our approach 
(see, e.g., Refs.~\cite{Gutsche:2013oea,Gutsche:2017wag}): 
$m_u = 0.2413$ GeV, $m_c = 1.6722$ GeV. 
The compositeness condition of Salam and Weinberg~\cite{Salam1962} 
gives one constraint equation between the coupling 
factors $g_{\Xi_{cc}^{++}},\,g_{\Sigma_c^{++}}$ and the size parameters
$\Lambda_{\Xi_{cc}^{++}},\,\Lambda_{\Sigma_c^{++}}$
charactering the nonlocal distribution 
$F_{\Xi_{cc}^{++}}(x;x_1,x_2,x_3),\,F_{\Sigma_c^{++}}(x;x_1,x_2,x_3)$,
respectively. 
As size parameter we use 
$\Lambda_{\Sigma_c^{++}}=0.867$ GeV (unified size parameter  
for the $J^P = \frac{1}{2}^+$ single charm baryons 
fixed in Refs.~\cite{Gutsche:2013oea}) and 
consider $\Lambda_{\Xi_{cc}^{++}}$ as a free parameter. 
The size parameters of light and heavy baryons in our approach 
are varied in the region $0.5-1$~GeV. Therefore, 
in our calculations we will vary $\Lambda_{\Xi_{cc}^{++}}$ in 
the interval $0.5-1$~GeV. We found that the results for the decay 
widths $\Gamma(\Xi_{cc}^{++} \to \Sigma_c^{++} + \bar K^{*0} (\bar K^0))$ 
are very stable in this region of the size parameter 
$\Lambda_{\Xi_{cc}^{++}}$. In the following discussion 
we will indicate the dependence of our results on the choice of
$\Lambda_{\Xi_{cc}^{++}}$ 
in the interval $0.5-1$~GeV in the form $A \pm \Delta A$. 

The leptonic decay constants $f_{K^*} = 212$ MeV and $f_K = 161.3$ MeV 
evaluated in our approach are in good agreement with data: 
$f_{K^*} = (217 \pm 7)$ MeV and 
$f_{K} = (156.1 \pm 0.8)$ MeV~\cite{PDG:2016}.  

Let us now list our numerical results for the four helicity amplitudes.
They are
    \be
\begin{array}{lcrlcr}
H_{\frac12 0}  &=&    3.0 \pm 0.1\, {\rm GeV}\,, 
\qquad & \qquad 
H_{-\frac12 0} &=& - (9.6 \pm 0.4)\, {\rm GeV}\,,  
\\[1.1ex]
H_{\frac12 1}  &=& - (3.0 \pm 0.3)\, {\rm GeV}\,,  
\qquad & \qquad 
H_{-\frac12 -1}&=&   10.0 \pm 0.5\, {\rm GeV} \,.
\\
\end{array}
\en
For the sum of the moduli squared of the helicity amplitudes one
obtains 
\bea
 {\cal H}_N &=& {\cal H}_U + {\cal H}_L 
             =  210.2 \pm 17.9 \,\,{\rm GeV}^2\,,\nonumber\\
 {\cal H}_U &=& 109.0 \pm 10.2 \,\,{\rm GeV}^2\,,\nonumber\\
 {\cal H}_L &=& 101.2 \pm  7.7 \,\,{\rm GeV}^2\,,
\ena
which leads to the partial decay width  
\be
\Gamma(\Xi_{cc}^{++} \to \Sigma_c^{++}\, + \,\bar K^{* 0}) 
= (0.21 \pm 0.02) \times 10^{12} \, {\rm s}^{-1}\,.
\en
We have also analyzed the decay
$\Xi_{cc}^{++} \to \Sigma_c^{++}\, + \,\bar K^{0}$ using the same dynamics as
for the decay $\Xi_{cc}^{++} \to \Sigma_c^{++}\, + \,\bar K^{* 0}$. We obtain
\be
\Gamma(\Xi_{cc}^{++} \to \Sigma_c^{++}\, + \,\bar K^{0}) 
= (0.05 \pm 0.01) \times 10^{12} \, {\rm s}^{-1}\,. 
\en 
The $\bar K^{*0}$ mode is about four times stronger than the one
including the $\bar K^0$. 
In order to convert the partial rate into a branching ratio one would
need the total width or, equivalently, the lifetime value of the 
$\Xi_{cc}^{++}$.
Neither of these are known experimentally. There have been several attempts to
calculate the lifetime of the $\Xi_{cc}^{++}$ based on the optical theorem
for the inclusive decay width combined with the Operator Product Expansion 
for the transition currents together with a heavy quark mass expansion. The
results are in the range
of $430\,{\rm fs}\,- 670 \,{\rm fs}$~\cite{Kiselev:1998sy,Chang:2007xa}. 
As a median value we take $\tau_{\Xi_{cc}^{++}}= 500\,{\rm fs}$. 
For the branching ratios we obtain
\bea
B(\Xi_{cc}^{++} \to \Sigma_c^{++}\, + \,\bar K^{* 0})
&=&\left(\frac{\tau_{\Xi_{cc}^{++}}}{500\,{\rm fs}}\right) 
\cdot (10.5 \pm 1)\,\%\,, 
\nonumber\\
B(\Xi_{cc}^{++} \to \Sigma_c^{++}\, + \,\bar K^{0})&=&
\left(\frac{\tau_{\Xi_{cc}^{++}}}{500\,{\rm fs}}\right) 
\cdot (2.5 \pm 0.5)\,\%\,.  
\nonumber
\ena

\section{Polarization, longitudinal/transverse helicity fraction and 
angular decay distributions}

We treat the decaying $\Xi_{cc}^{++}$ as being unpolarized. In principle, the
$\Xi_{cc}^{++}$ could acquire a nonzero transverse polarization in the hadronic
production process. However, since one is averaging over the rapidities
of the production process the  $\Xi_{cc}^{++}$ is effectively unpolarized
(for more details see~\cite{Gutsche:2017wag}). The baryon-side decay
$\Sigma_c^{++} \to \Lambda_c^+\ \pi^+$
is a strong decay and, even though the $\Sigma_c^{++}$ is polarized, the
decay $\Sigma_c^{++} \to \Lambda_c^+\ \pi^+$ possesses zero analyzing power
to resolve the polarization of the $\Sigma_c^{++}$, i.e. the azimuthal
angle and the helicity angle decay distribution of the decay 
$\Sigma_c^{++} \to \Lambda_c^+\pi^+$ is uniform. 
For the meson-side decay $\bar K^{*0} \to K^- \pi^+$ one obtains the angular
decay distribution
\bea
\label{angdis1}
\frac{d\,\Gamma(\Xi_{cc}^{++} \to \Sigma_c^{++} 
  +\bar K^{*0} (\to K^-  \pi^+))}{d\cos\theta_V} 
&=& B(\bar K^{*0} \to K^-  \pi^+)\frac{G_F^2}{32 \pi} \, \frac{|{\bf p_2}|}{M_1^2} \, 
|V_{ij} V^\ast_{kl}|^2 \, C_{\rm eff}^2 \, f_V^2 \, M_V^2  \,\,{\cal H}_N  \nn
&\times&\bigg(\frac 32 \cos^2\theta_V \,{\cal F}_L 
+\frac 34 \sin^2\theta_V \,{\cal F}_T \bigg) 
\ena
where $B(\bar K^{*0} \to K^-  \pi^+)=2/3$ is the branching ratio of the
decay $\bar K^{*0} \to K^-  \pi^+$.
The angular decay distribution~(\ref{angdis1}) involves the helicity
fractions of the $\bar K^{* 0}$ defined by
\be
   {\cal F}_L = \frac{|H_{\frac12 0}|^2 +|H_{-\frac12 0}|^2}{{\cal H}_N}
   = 0.48 \pm 0.01\,, \qquad
   {\cal F}_T = \frac{|H_{\frac12 1}|^2 +|H_{-\frac12 -1}|^2}{{\cal H}_N} 
   = 0.52 \pm 0.01 \,.   
   \en
   This has to be compared to the unpolarized case 
   ${\cal F}_L=1/3$ and ${\cal F}_T = 2/3$ which is e.g. realized at the
   zero recoil point $q^2=(M_1-M_2)^2$ where there is only the axial vector
   $S$-wave excitation of the final ($\Sigma_c^{++}\,\bar K^{*0}$)--state
   with $\sqrt{2} H^A_{1/2\, 0} = H^A_{1/2\, 1}$
   (``allowed Fermi--Teller transition'').
   Our results for the helicity fractions considerably deviate from
   their unpolarized values leading to a pronounced
   $\cos\theta_V$--dependence of the angular decay distribution~(\ref{angdis1})
   which is quite close to $W(\theta_V)\sim 3/8 (1+\cos^2\theta_V)$.

   The longitudinal polarization of the daughter baryon $\Sigma_c^{++}$ 
   depends on the polar emission angle $\theta_V$ via
   \be
   \label{polsigma}
   P_{\Sigma_c^{++}}(\cos\theta_V)=\frac{
    \frac 34 \sin^2\theta_V \Big(|H_{\frac12 1}|^2 -|H_{-\frac12 -1}|^2\Big)
  + \frac 32 \cos^2\theta_V \Big(|H_{\frac12 0}|^2 -|H_{-\frac12 0}|^2 \Big)}
   {\frac 34 \sin^2\theta_V \Big(|H_{\frac12 1}|^2 +|H_{-\frac12 -1}|^2\Big)
   +\frac 32 \cos^2\theta_V \Big(|H_{\frac12 0}|^2 +|H_{-\frac12 0}|^2 \Big)} 
\,.
\en
When averaged over $\cos\theta_V$ (one has to integrate the numerator and
denominator separately) one has  
\be
   P_{\Sigma_c^{++}}=\frac{
    \Big(|H_{\frac12 1}|^2 -|H_{-\frac12 -1}|^2 \Big)
   + \Big(|H_{\frac12 0}|^2 -|H_{-\frac12 0}|^2 \Big)}
   {{\cal H}_N} = -(0.83 \pm 0.01) \,.    
     \en
     As mentioned before the polarization of the $\Sigma_c^{++}$ is not
     measurable in its strong decays. However, the $\Sigma_c^{++}$
     transfers its polarization to the $\Lambda_c^+$ in the strong decay 
     $\Sigma_c^{++} \to \Lambda_c^+\ \pi^+$.      
     The average longitudinal polarization of the $\Lambda_c^+$ can be
     calculated to be (we average over $\cos\theta_V$):
     \be
     P_{\Lambda_c^+}(\theta_B)=
     \frac{|H_{\frac12 0}|^2 - |H_{-\frac12 0}|^2 
         + |H_{\frac12 1}|^2 - |H_{-\frac12 -1}|^2}
       {{\cal H}_N}\,\,\cos\theta_B\,=\,-(0.83 \pm 0.01) \,\cos\theta_B
       \en
       where $\theta_B$ is the angle between the direction of the $\Lambda_c^+$
       and the original flight direction of the $\Sigma_c^{++}$, all in the
       rest frame of the $\Sigma_c^{++}$.

       For the decay $\Xi_{cc}^{++} \to \Sigma_c^{++} + \bar K^{0}$ 
       we find a slightly larger value of the longitudinal polarization 
       of the $\Sigma_c^{++}$ given by
       \be
       P_{\Sigma_c^{++}}(\Xi_{cc}^{++} \to \Sigma_c^{++} 
       +\bar K^{0})=\frac{
       |H_{\frac12 t}|^2 -|H_{-\frac12 t}|^2}
       {{\cal H}_S} = -(0.95 \pm 0.02) \,.    
       \en
       In principle, the polarization of the $\Lambda_c^+$ 
       can be analyzed in its
       weak decay $\Lambda_c^+ \to pK^-\pi^+$. For example, one could attempt
       to measure nonvanishing values of the expectation value
       $\la \cos\theta_i \ra$ where $\theta_i$ is the polar angle
       between the polarization direction of the $\Lambda_c^+$ 
       and either one of the
       three decay particles ($i=p,K^+,\pi^-$) or the normal of the decay
       plain (see an exemplary analysis of a weak $(1 \to 3)$--particle
       decay in e.g.~\cite{Korner:1998nc}). To our knowledge the weak decay
       $\Lambda_c^+ \to pK^-\pi^+$ has not been completely calculated yet
       except for an analysis of the subchannels
       $\Lambda_c^+ \to p \bar K^{*0}$ and 
       $\Lambda_c^+ \to \Delta^{++}K^-$~\cite{Konig:1993wz}.
       
\section{Summary and conclusion}

We have discussed in some detail the possibility that 
the new double charm state found in the invariant mass distribution of 
$(\Lambda_c^+\,K^-\,\pi^+\,\pi^+)$ can be attributed to the decay chain 
$\Xi_{cc}^{++} \to \Sigma_c^{++} (\to \Lambda_c^+ \pi^+) +\bar K^{*0} 
(\to K^-  \pi^+)$. The hypothesis can be tested experimentally by looking
at the decay distributions of the particles involved in the cascade
decay. For once one can check whether there are significant peaks at the
$ \Sigma_c^{++}$ and $\bar K^{*0}$ masses in the ($\Lambda_c^+ \pi^+$) and
($K^-  \pi^+$) invariant mass distributions, respectively. If there is a
significant continuum background one would have to place relevant cuts
on the invariant mass distribution to obtain the appropriate cascade decay
channels discussed in this paper. One can then go on
and check on the angular decay distributions in the respective cascade decays
which have been written down in this paper. We have also discussed the
decay $\Xi_{cc}^{++} \to \Sigma_c^{++} + \bar K^{0}$ which we predict to have 
a branching ratio four times smaller than that of the decay
$\Xi_{cc}^{++} \to \Sigma_c^{++} + \bar K^{*0}$. It would nevertheless be
interesting to experimentally search for this decay mode.

It would also be worthwhile to experimentally check on further nonleptonic 
decay channels of the double charm state $\Xi_{cc}^{++}$ (see also 
Refs.~\cite{Li:2017ndo}). 
For once there are the decay channels
$\Xi_{cc}^{++} \to \Xi_c^{* +}(\to \Xi_c^0 +\pi^+)+ \pi^+(\rho^+)$ and
$\Xi_{cc}^{++} \to \Xi_c^{* ++}(\to \Xi_c^+ +\pi^+)+ \rho^0$. Experimentally
more challenging would be the decay channels
$\Xi_{cc}^{++} \to \Xi_c^{'+} (\to \Xi_c^+ +\gamma )+ \pi^+(\rho^+)$,
 and
$\Xi_{cc}^{++} \to \Xi_c^{* ++}
 (\to \Xi_c^+ +\pi^+)+ \pi^0$ because their detection would require photon
 identification. The above two-body nonleptonic
decay modes belong to the same class of processes as the decays
$\Xi_{cc}^{++} \to \Sigma_c^{++} + \bar K^{*0} (\bar K^0)$ 
in that they are solely 
contributed to by the factorizing (or tree graph) contribution.
A possible $W$-exchange contribution (color commensurate ``C'' in the
terminology of~\cite{Leibovich:2003tw}) is forbidden by 
the K\"orner, Pati, Woo theorem~\cite{Korner:1970xq}. 
The calculation of the above rates
proceeds in the same way as the calculation in this paper and
will be the subject of a future publication~\cite{Gutsche:2017tb}.

\begin{acknowledgments}

This work was funded 
by the German Bundesministerium f\"ur Bildung und Forschung (BMBF)
under Project 05P2015 - ALICE at High Rate (BMBF-FSP 202):
``Jet- and fragmentation processes at ALICE and the parton structure     
of nuclei and structure of heavy hadrons'',
by CONICYT (Chile) PIA/Basal FB0821, by Tomsk State University Competitiveness
Improvement Program and the Russian Federation program ``Nauka''
(Contract No. 0.1764.GZB.2017). 
The research is carried out at Tomsk Polytechnic University within 
the framework
of Tomsk Polytechnic University Competitiveness Enhancement Program grant. 
M.A.I.\ acknowledges the support from  PRISMA Cluster of Excellence
(Mainz Uni.). M.A.I. and J.G.K. thank the Heisenberg-Landau Grant for
partial support.

\end{acknowledgments}

\end{document}